\documentclass[useAMS,usenatbib,letter]{mnras}
\usepackage[dvips]{graphicx}
\usepackage{hyperref}
\usepackage{epstopdf}
\usepackage{times}
\usepackage{amsmath}
\usepackage{amssymb}
\usepackage{color}
\hypersetup{
 colorlinks,
 citecolor=blue,
 linkcolor=blue
}
\title[Binary star spatial mixing]{Spatial mixing of binary stars in multiple-population globular clusters}
\author[J. Hong et al.] {Jongsuk Hong$^{1,2}$\thanks{E-mail: jongsuk.hong@pku.edu.cn (JH)}, Saahil Patel$^{2,3}$, Enrico Vesperini$^2$, Jeremy J. Webb$^{2,4}$ \newauthor and Emanuele Dalessandro$^5$
\\
 $^1$ Kavli Institute for Astronomy and Astrophysics, Peking University, Yi He Yuan Lu 5, HaiDian District, Beijing 100871, China\\
 $^2$ Department of Astronomy, Indiana University, Swain West, 727 E. 3rd Street, Bloomington, IN 47405, USA\\
 $^3$ College of Electrical Engineering \& Computer Science, Syracuse University, 111 College Pl, Syracuse, NY 13210, USA\\
 $^4$ Department of Astronomy and Astrophysics, University of Toronto, 50 St. George Street, Toronto ON M5S 3H4, Canada\\
 $^5$ INAF-Astrophysics and Space Science Observatory, Via Gobetti 93/3, I-40129 Bologna, Italy
}
\begin{document}

\newcommand{\srC}{{MPr5f1}}
\newcommand{\xg}{x_{\rm g,0} }
\newcommand{\rratio}{R_{\rm h,FG}/R_{\rm h,SG}}
\newcommand{\rh}{R_{\rm h}}
\newcommand{\rhfg}{R_{\rm h,FG}}
\newcommand{\rhsg}{R_{\rm h,SG}}
\newcommand{\trh}{\tau_{\rm rh,0}}

\date{Accepted 2018 December 3. Received 2018 November 29; in original form 2018 October 30}
\maketitle

\label{firstpage}

\begin{abstract}
 We present the results of a study aimed at investigating the effects of dynamical evolution on the spatial distribution and mixing of primordial binary stars in multiple-population globular clusters.
 Multiple stellar population formation models predict that second-generation (SG) stars form segregated in the inner regions of a more extended first-generation (FG) cluster. 
Our study, based on the results of a survey of $N$-body simulations, shows that the spatial mixing process for binary stars is more complex than that of single stars since additional processes such as binary ionization, recoil and ejection following binary-single and binary-binary interactions play a key role in determining the spatial distribution of the population of surviving binaries. The efficiency and relative importance of these additional effects depends on the binary binding energy and determines the timescale of the spatial mixing of FG and SG binaries. Our simulations illustrate the role of ionization, recoil and ejection combined with the effects of mass segregation driven by two-body relaxation and show that the complex interplay of all these processes results in a significant extension of the time needed for the complete spatial mixing of FG and SG binaries compared to that of single stars. Clusters in which FG and SG single stars have already reached complete spatial mixing might be characterized by a significant radial gradient in the ratio of the FG-to-SG binary fraction. The implications of the delayed mixing of FG and SG binaries for the differences between the kinematics of the two populations are discussed.
\end{abstract}


\section{Introduction}
\label{sec:intro}

Many spectroscopic and photometric studies \citep[see e.g.][and references therein]{2012A&ARv..20...50G,2015AJ....149...91P,2017MNRAS.464.3636M} have revealed the presence of two or more stellar populations with different chemical compositions in globular clusters.

A variety of models \citep[see e.g.][]{2007A&A...464.1029D,2008MNRAS.391..825D,2012MNRAS.423.1521D,2009A&A...507L...1D,2013MNRAS.436.2398B,2016MNRAS.458.2122D,2018MNRAS.478.2461G,2018MNRAS.tmpL.171B} have been proposed in the literature for the origin of multiple stellar populations and the sources of gas out of which second-generation stars (hereafter SG) might have formed (in some models stars with different chemical compositions belong to the same generation; hereafter for convenience, however, we indicate stars that formed from chemically processed gas as SG); despite the significant differences in many of the fundamental ingredients of the different models, they all agree that SG stars should form more concentrated in the cluster's inner regions.
A number of observational studies have confirmed that in some clusters SG stars are indeed more concentrated than FG stars and still preserve some memory of the initial spatial trend predicted by theoretical models \citep[e.g.][]{2007ApJ...654..915S,2009A&A...507.1393B,2011A&A...525A.114L,2012ApJ...744...58M,2014ApJ...780...94C,2016MNRAS.458.4162M,2016MNRAS.463..449S}.
In other clusters \citep[see e.g.][]{2013A&A...555A.143M,2014ApJ...791L...4D,2017ApJ...844...77L,2018MNRAS.477.2004N}, FG and SG stars are spatially mixed and any initial trend must have been erased by the effects of the clusters' long-term dynamical evolution.

A few studies of the dynamics of the spatial mixing process \citep{2013MNRAS.429.1913V,2018MNRAS.476.2731V,2015MNRAS.454.2166M} have illustrated the role of internal two-body relaxation and mass loss in erasing initial differences in the spatial distributions of multiple populations and shown that a cluster must lose (due to the effects of two-body relaxation) at least 60-70 per cent of its initial mass
before reaching complete mixing of all the populations \citep[see also][for studies of mixing of stars with different helium abundances]{2018ApJ...859...15D,2018MNRAS.481.3027F}.

The formation and dynamical evolution of multiple populations process also can lead to differences in their kinematic properties \citep{2011MNRAS.412.2241B,2015MNRAS.450.1164H,2015ApJ...810L..13B} and some observational studies have indeed revealed differences in the rotation and anisotropy in the velocity dispersion of FG and SG stars \citep{2013ApJ...771L..15R,2015ApJ...810L..13B,2017MNRAS.465.3515C,2018MNRAS.479.5005M,2018ApJ...861...99L}.

Another manifestation of the dynamics of multiple-population clusters and of the different structural scales of the two populations appears in the evolution of primordial binaries.
Numerous studies have explored the dynamical evolution of binary stars in star clusters and shown the importance of binary stars for the structural evolution of star clusters
 \citep[e.g.][]{2006MNRAS.368..677H,2007ApJ...665..707H,2007MNRAS.374..344T,2010ApJ...719..915C,2013ApJ...765....4D,2013ApJ...779...30G,2015ApJ...805...11G,2013MNRAS.436.1497L,2015MNRAS.446..226L}.

The internal (e.g. binding energy, mass ratio) and external properties (spatial distribution and kinematics) of a population of binary stars are determined by the effects of two-body relaxation and the possible outcomes of binary-binary and binary-single interactions. These interactions can result in the ionization of the interacting binary, in the evolution of the binaries' internal binding energy and induce a significant recoil velocity possibly causing the binary ejection \citep[see e.g.][]{2003gmbp.book.....H}.

These interactions can be very important also for the formation of exotic populations such as blue stragglers, cataclysmic variables, milli-second pulsars, and low-mass X-ray binaries. A number of observational studies and theoretical investigations based on realistic numerical simulations have studied the possible link between the properties and number of exotic stellar populations and the dynamical properties of globular clusters \citep[see e.g.][]{2003ApJ...591L.131P,2006ApJ...646L.143P,2010ApJ...714.1149H,2011MNRAS.416.1410L,2013MNRAS.428..897L,2012Natur.492..393F,2013ApJ...777..106C,2017MNRAS.466..320H,2017MNRAS.464.2511H}.

In the context of the study of the binary dynamics of multiple-population clusters, \citet{2011MNRAS.416..355V} and \citet[][hereafter Papers I and II]{2015MNRAS.449..629H,2016MNRAS.457.4507H} have shown that the differences between FG and SG spatial distributions lead to a preferential disruption and ejection of SG primordial binaries resulting in a larger global FG binary fraction. This trend is consistent with the findings of a few observational studies \citep{2010ApJ...719L.213D,2015A&A...584A..52L,2018ApJ...864...33D} that have revealed a larger fraction of FG binaries, in general agreement with our theoretical predictions.

In this paper we study how the spatial mixing process of FG and SG binaries is affected by the additional dynamical processes (ionization, recoil and ejection following binary-binary and binary-single interactions) determining the evolution and survival of binary stars. 
The outline of this paper is the following. In section 2, we introduce our methods and the initial conditions. We present our results in section 3. Our conclusions follow in section 4. 

\section{Methods and Initial Conditions}
The results of this paper are based on the $N$-body simulations presented in \hyperlink{P1}{Papers I} and \hyperlink{P2}{II}.
We summarize here the main properties of the simulations and we refer to those papers for further details.

The simulations were run with the GPU-accelerated version of the code {\sc nbody6-GPU} \citep{2003gnbs.book.....A,2012MNRAS.424..545N} running on the {\sc big red ii} GPU cluster at Indiana University.

In the models explored the FG and SG subsystems have initially the same total mass and they both follow a \citet{1966AJ.....71...64K} density profile with central dimensionless potential
$W_0=7$. However, the SG subsystem is initially centrally concentrated in the inner regions of the FG system as suggested by a number of multiple population formation models. We have selected models from \hyperlink{P1}{Papers I} and \hyperlink{P2}{II} in which the initial ratio of the FG to the SG half-mass radius ($R_{\rm h,FG}/R_{\rm h,SG}$) is equal to 5.
All of the simulations start with equal-mass particles and we do not consider the effects of stellar evolution. We focus our attention on the effects of two-body relaxation and mass loss through the cluster's tidal limit set by the external tidal field.
The effects of the host galaxy tidal field are included assuming a simple point mass model for the host galaxy; stars moving beyond a radius equal to two times the tidal radius are removed from the simulation. We assumed that the cluster is initially tidally truncated.

In this study, we fixed the initial binary fraction, $f_{\rm b,0}=N_{\rm bin}/(N_{\rm s}+N_{\rm bin})$ equal to 0.10 (where $N_{\rm s}$ is the number of single stars and $N_{\rm bin}$ is the number of binaries; $N=N_{\rm bin}+N_{\rm s}$; thus, $N_{\rm tot}=N_{\rm s}+2N_{\rm bin}$ where $N_{\rm tot}$ is the total number of particles). 
The initial number of single and binary stars is $N=20,000$ where $N_{\rm tot}=22,000$ for simulations with $f_{\rm b,0}=0.1$. 
We assume that FG and SG binaries have the same initial binding energy distribution. 
In this paper we considered binaries with initial global hardness parameters, $\xg$, ranging from 3 to 800; $\xg$ is defined as $E_{\rm b}/(m \sigma_{\rm SP}^2)$ where $E_{\rm b}$ is the absolute value of the binary binding energy, and $\sigma_{\rm SP}$ the 1-D velocity dispersion of all stars in a reference single-population system (see \hyperlink{P1}{Papers I} and \hyperlink{P2}{II} for more details). 

\begin{table}
 \begin{center}
 \caption{Initial parameters for selected models. For all models $N=20,000$ where $N=N_{\rm s}+N_{\rm bin}$ is the total number of single, $N_{\rm s}$, and binary, $N_{\rm bin}$, particles. For all models the initial binary fraction  $f_{\rm b,0}=N_{\rm bin}/(N_{\rm s}+N_{\rm bin})$ is equal to 0.1. 
For all models the initial ratio of the FG to the SG initial half-mass radii, $R_{\rm h,FG}/R_{\rm h,SG}$, is equal to 5. $\xg$ initial hardness parameter (see section 2 for definition).}
 \begin{tabular}{l c}
 \hline
 \hline
 Model id. & $x_{\rm g,0}$ \\
 \hline
 \srC{x3} & 3\\
 \srC{x20}  & 20\\
 \srC{x3-20} & 3-20$^a$\\
 \srC{x50}  & 50\\
 \srC{x800} & 800\\
 \srC{x3-800}  & 3-800$^a$\\
 \hline
 \srC{m2}  & N/A$^b$\\
 \hline
 \label{tbl1}
 \end{tabular}
 \end{center}
\begin{flushleft} 
$^a$ A uniform distribution in binding energy is assumed.\\
$^b$ Binaries replaced by single stars with mass equal to 2$m_{\rm s}$.\\
\end{flushleft}
\end{table}
In order to better illustrate the dependence of the binary spatial evolution on the hardness parameter, we also present the results for simulations including binaries with a single value of $\xg$=3, 20, 50, 800.
We have also selected two simulation models including binaries with a uniform binding energy distribution from $\xg=3$ to $\xg=20$ and from $\xg=3$ to $\xg=800$.
We performed an additional simulation in which binaries are replaced by single stars with 2$m_{\rm s}$ (\srC m2), in order to disentangle the effects on the spatial mixing due to segregation from those due to the other processes affecting binary stars (ionization,recoil, ejection).

We summarize in Table \ref{tbl1} the main properties of the models presented in this paper.

\section{Results}
\subsection{Spatial distribution and mixing}
\begin{figure}
 \includegraphics[width=85mm,trim={0mm 4mm 0mm 0mm}]{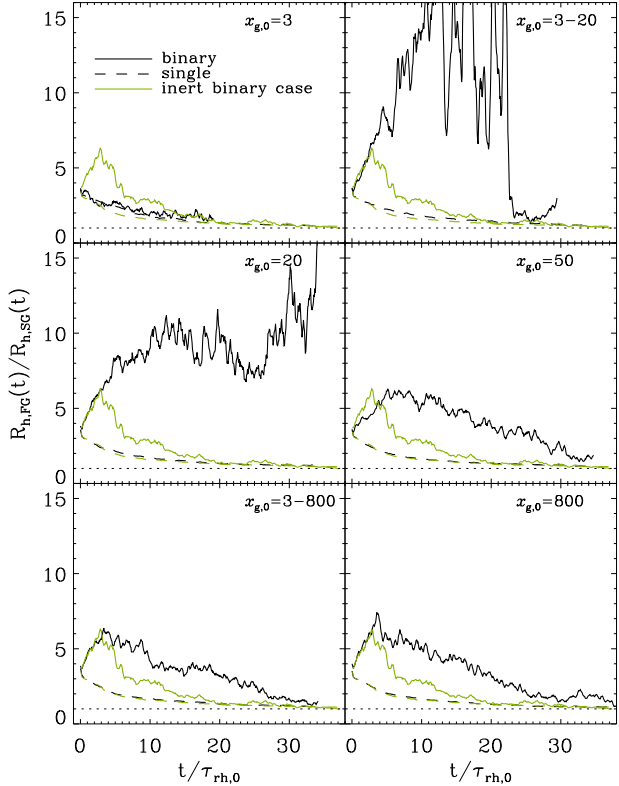}
 \caption{Time evolution of the ratio of the half-mass radii of FG stars to SG stars. Models with $\xg=$ 3, 3-20, 20, 50, 3-800 and 800 are shown in different panels. 
 The solid and dashed lines represent the ratio of half-mass radii for binaries and single stars, respectively. For the inert binary model (green lines), stars with 2$m$ are plotted for comparison with the evolution of binaries.
 The horizontal dotted line at $\rratio=1$ indicates the completion of the spatial mixing.}\label{F1}
\end{figure}
\begin{figure}
 \includegraphics[width=85mm,trim={0mm 4mm 0mm 0mm}]{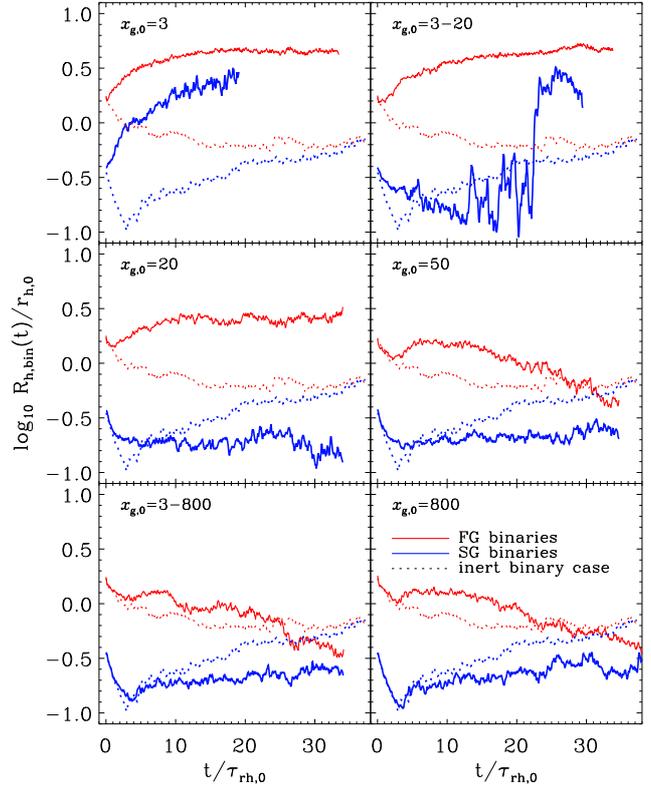}
 \caption{Time evolution of the half-mass radii of both FG and SG binary populations. The same models as in Fig. \ref{F1} are presented. Solid lines show the evolution for FG and SG binaries. The FG and SG inert binaries are plotted with dotted lines for comparison.
}\label{F2}
\end{figure}
\begin{figure*}
 \includegraphics[width=178mm,trim={0mm 4mm 0mm 2mm}]{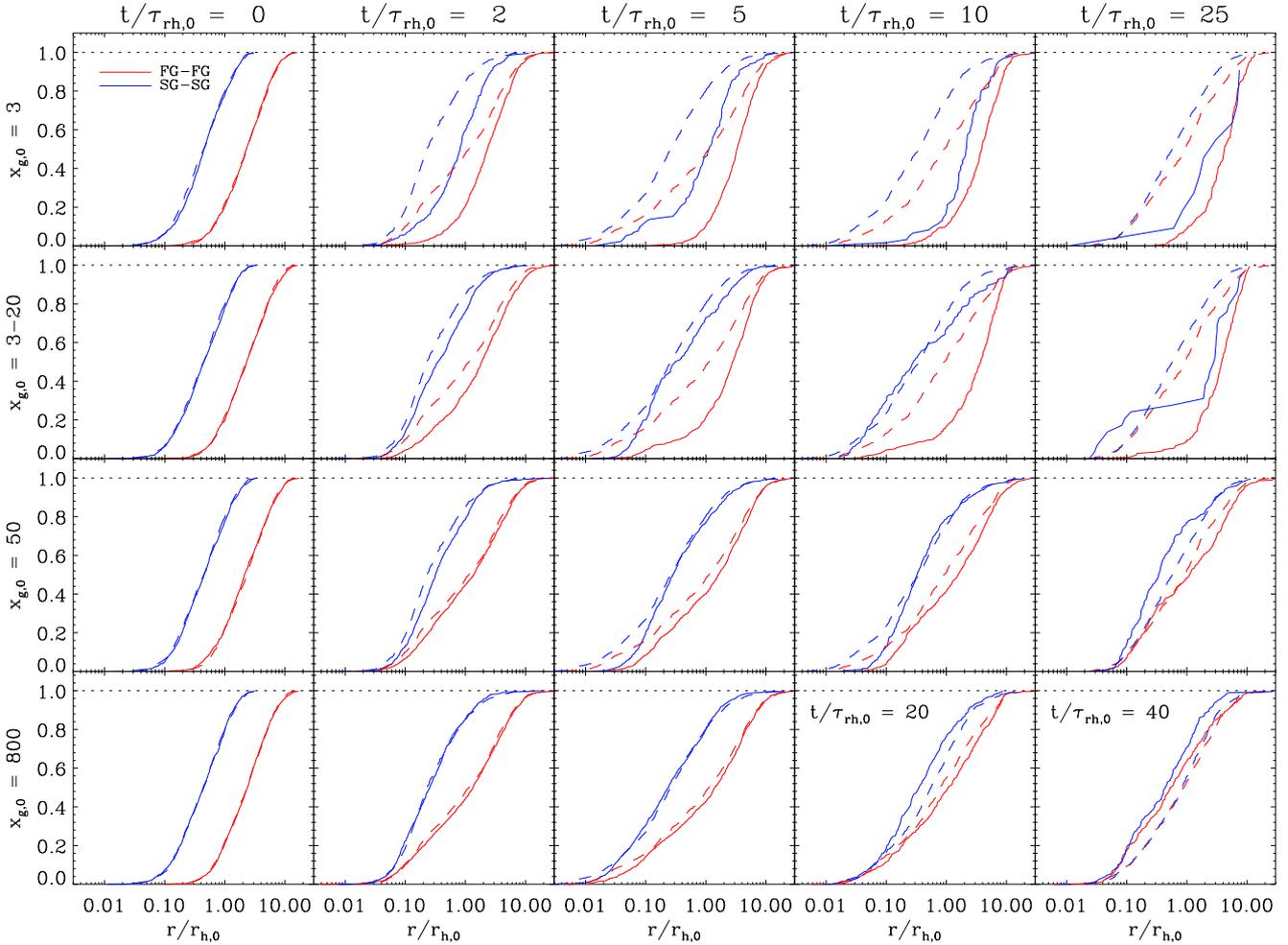}
 \caption{Cumulative radial distribution of FG (red lines) and SG (blue lines) binaries. From left to right, each panel shows the evolution of radial distribution. 
 Three different models with $\xg=3$, 3-20, 50 and 800 are presented (top to bottom). The cumulative radial distribution of inert binaries are shown as dashed lines for comparison.
 The last panel for the model with $\xg=800$, the distribution at $t\sim$40$\tau_{\rm rh,0}$ is shown to illustrate a complete spatial mixing of very compact binaries. 
 }\label{F3}
\end{figure*}
We start our analysis of the spatial mixing of FG and SG binaries by focusing our attention on the evolution of the half-mass radii of the two populations of binaries. Fig. \ref{F1} shows the time evolution of the ratio of the FG to the SG half-mass radius, $\rratio$, for single stars and for binaries (or inert binaries for the simulation \srC{}m2). The time evolution of half-mass radii for FG and SG binaries are shown in Fig. \ref{F2} separately.

These figures show several interesting results. First, in all the simulations, with the exception of the model starting with the softest binaries (\srC{}x3), $\rratio$ for binaries initially increases indicating that a growth in the differences between the spatial distribution of FG and SG binaries. The same effect has been found by \citet{2018MNRAS.476.2731V} in a study of the the spatial mixing of multi-mass systems in which they showed that the $\rratio$ for the most massive stars initially increases. As discussed in \citet{2018MNRAS.476.2731V}, this is a consequence of the different structures and timescales of the FG and SG subsystems: the SG subsystem is initially more compact and characterized by a shorter relaxation timescale and this implies that the process of mass segregation is more rapid for the SG massive stars than for the FG ones. SG massive stars therefore sink toward the cluster inner regions and their half-mass radius contracts more rapidly than that of FG stars leading to the observed time evolution of $\rratio$. Later in the evolution, after a few initial half-mass relaxation timescales, the evolution of $\rratio$ starts to decrease as two populations evolve toward spatial mixing.

In the case of the simulations presented in this paper, the dynamics is further enriched by the additional processes affecting binary stars (ionization, recoil, and ejection), their survival and the evolution of their spatial distribution.

As shown in Fig. \ref{F1}, although the initial evolution of $\rratio$ is the same for binaries and inert binaries, the long term evolution of $\rratio$ shows that binaries are characterized by a much longer mixing timescale than inert binaries and single stars. The results presented in this figure suggest that even clusters where single stars have reached complete spatial mixing may be characterized by a significant radial gradient in the fraction of FG and SG binaries.

The significant difference between the mixing time of binaries and single stars is due to the role of the additional processes affecting the survival and spatial distribution of all binaries. In the cluster's inner regions, for example, the frequent interactions with other single and binary stars may result in the disruption of the interacting binary, its ejection to the cluster's outermost regions or even its escape from cluster if the recoil speed is larger than the cluster's escape speed. 
The effect of these processes is therefore determined by the timescale of mass segregation which drives the binaries toward the cluster's inner regions where they can undergo more frequent interactions.

Finally, the relative importance of binary ionization and ejection depends on the binary's binding energy: the evolution of the spatial distribution of softer binaries will be dominated by the ionization process while for the harder binaries recoil/ejection will be the dominant process driving the evolution of their spatial distribution.

In general the evolution of the binary spatial distribution is driven by the interplay between mass segregation, which tends to make the binary's spatial distribution more concentrated, and ionization and recoil which preferentially remove binaries from the cluster's inner regions and therefore tend to make the binary's spatial distribution less concentrated. 

The results for simulations with binaries of different hardness illustrate the implications of the variation in relative importance of the various dynamical processes.
For example, for the simulations with the softest binaries considered in our survey, $\xg=3$, the evolution of $\rratio$ for binaries appears similar to that for single stars. 
However, the dynamics behind the evolution of $\rratio$ for these soft binary stars is different from that of single stars and is mainly driven by the rapid disruption of SG binaries in the cluster's inner and intermediate regions where the local value of the binary hardness factor is smaller than 1 (Fig. \ref{F2}; see also figure 6 of \hyperlink{P1}{Paper I}). The disruption of these soft binaries does not require an early phase of mass segregation and erases the spatial differences between the surviving binaries without any initial increase in the values of $\rratio$ due to mass segregation.

The effects of mass segregation are instead clearly visible in all the other simulations including harder binaries. Harder binaries can segregate to the central regions before being ionized by interactions with other single or binary stars or ejected from the cluster following these interactions. The more rapid segregation of SG binaries leads to initial increase in $\rratio$ as shown in Figs. \ref{F1} and \ref{F2}. The increase continues until the cluster reaches core collapse and the mass segregation process significantly slows down \citep[see e.g.][]{1996MNRAS.279.1037G,2007MNRAS.374..703K,2013MNRAS.430.2960H}.
The subsequent evolution is determined by the combined effects of binary disruption and binary segregation resulting in a significant delay of complete mixing compared to the inert binary case.

In order to further describe the overall effect of the different dynamical processes affecting the survival and evolution of binaries on the FG and SG binary mixing, we present in Fig. \ref{F3} the time evolution of the cumulative radial distribution of FG and SG binaries for four different models with different values of the initial hardness parameter of the binary population.

\begin{figure}
 \includegraphics[width=85mm,trim={8mm 8mm 2mm 3mm}]{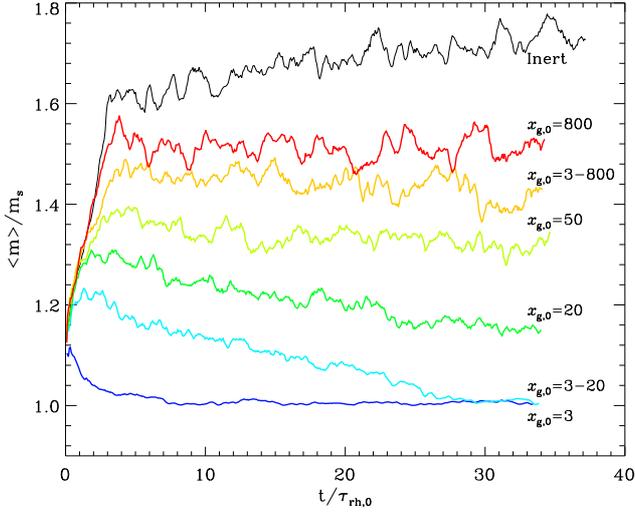}
 \caption{Time evolution of mean mass of stars (normalized to the mass of single stars $m_{\rm s}$) within 5\% Lagrangian radius. For this calculation binaries are considered as single stars with positions equal to those of their center-of-mass and masses equal to 2$m_{\rm s}$. 
 }\label{F4}
\end{figure}
The cumulative radial distribution of FG and SG inert binaries in the \srC{}m2 are also shown in order to illustrate the spatial mixing of stars with the same mass of binaries but not affected by other processes such as ionization and ejection due to the recoil velocity acquired during interactions with other stars.
At $t=0$, FG and SG binaries in all models have the same cumulative distribution. For the \srC{}x3 model, the rapid disruption of SG binaries significantly depletes the inner binary population as illustrated by the shift of the SG binary cumulative radial distribution towards the outer regions. For the \srC{}x3-20 model, the evolution of the distribution of FG binaries is similar to that for the \srC{}x3 model. However, the presence of harder binaries implies that the effects of ionization are not as strong as in the previous case and the evolution of the SG cumulative radial distribution is more similar to that of inert binaries.

It is interesting to note the development of an approximately flat portion in the cumulative radial distribution of SG binaries in the \srC{}x3 and x3-20 simulations. 
The formation of this shell depleted in SG binaries is due to the combined effect of binary segregation and disruption; binaries are rapidly disrupted in the inner regions but as the segregation time-scale increases with the distance from the cluster's centre. Therefore, there is no replenishment of binaries beyond the gap because the segregation time-scale is longer than the cluster's age (see also the figure 7 in \hyperlink{P1}{Paper I}).
The development of this gap is also responsible for the rapid variation in $\rhsg$ and $\rratio$ around $t\sim$23$\tau_{\rm rh,0}$ in Figs. \ref{F1} and \ref{F2} for the \srC{}x3-20. 

\begin{figure}
 \includegraphics[width=85mm,trim={0mm 6mm 0mm 0mm}]{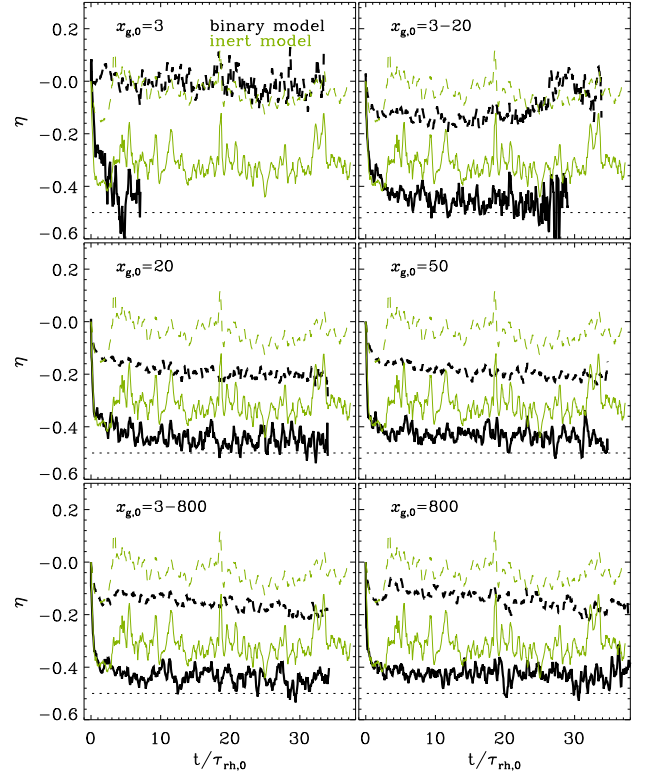}
 \caption{Time evolution of $\eta$ (see section \ref{sec:seg} for the definition)) within 5\% Lagrangian radius (solid lines) and half-mass radius (dashed lines). The horizontal dotted line indicates the complete equipartition ($\eta=-0.5$). Green lines represent the $\eta$ evolution of inert binary model, for comparison.
 }\label{F5}
\end{figure}
For the \srC{}x800 model, the cumulative distributions for FG and SG binaries are, in general, very similar to those in the inert binary model. Eventually, the spatial mixing of FG and SG binaries is achieved at $t\sim$40$\tau_{\rm rh,0}$.

Fig. \ref{F4} shows the time evolution of the mean mass of stars within the 5\% Lagrangian radius. For this calculation binaries are considered as single stars with positions given by their center-of-mass and masses equal to 2$m_{\rm s}$. 
Binary disruption leads to an increase of the number of star counts within the radius, while keeping the total mass unchanged. Thus, the mean mass of stars in the central region decreases as binaries are disrupted. 
Fig. \ref{F4} shows the competing effects of  binary segregation and disruption in the cluster's central regions. 
The evolution of mean mass in the inert binary model shows the effect of mass segregation without disruption.
For the case with softest binaries ($\xg=3$), the mean mass of stars decreases rapidly and converges to $m_{\rm s}$; in this case, as binaries segregate toward the central regions they are rapidly disrupted. There is no increase in the mean mass because binary disruption efficiently counteracts the effects of segregation. 

For the \srC{}{x3-20} model, the effect of disruption overcomes that of segregation at $t\sim$1.5$\tau_{\rm rh,0}$, and the mean mass decreases afterward. 
As the initial hardness of binaries increases, disruption becomes less important and the mean mass of stars in the cluster's inner regions reaches larger values.
In all models, a balance between the disruption or ejection of binaries and the replenishment of binaries due to the segregation from outer shells is eventually reached in different values of the mean mass.

\begin{figure}
 \includegraphics[width=84mm,trim={3mm 3mm 0mm 3mm}]{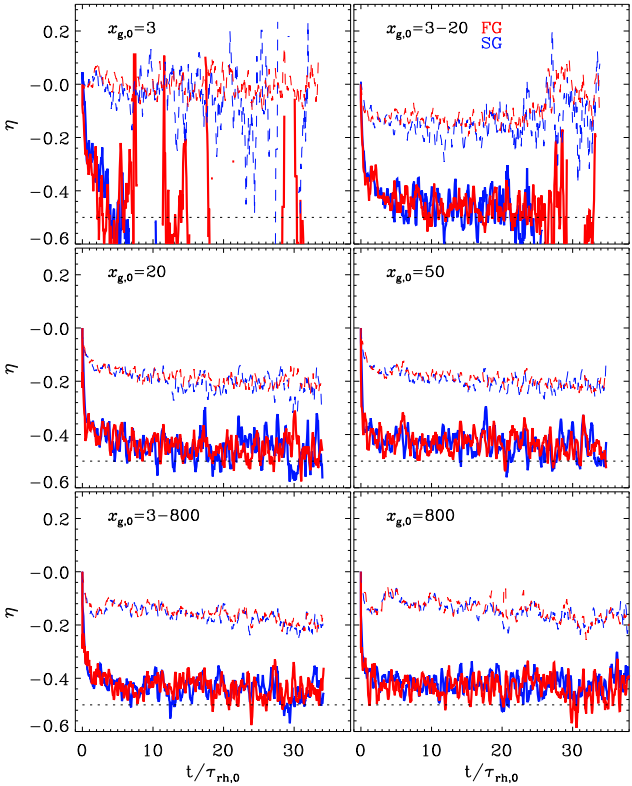}
 \caption{Same as Fig. \ref{F5} but for FG (red) and SG (blue) stars separately.
 }\label{F6}
\end{figure}
\subsection{Binary segregation and equipartition}\label{sec:seg}
\begin{figure}
 \includegraphics[width=84mm,trim={3mm 1mm 4mm 4mm}]{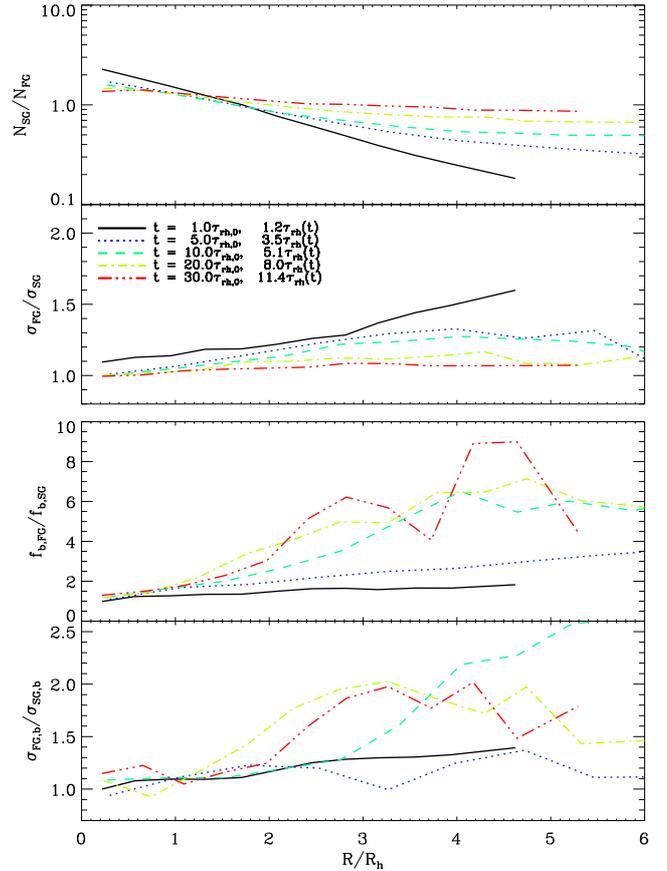}
 \caption{Time evolution of radial distributions of the SG-to-FG number ratio (upper panel), the FG-to-SG ratio of velocity dispersion with binaries as a center-of-mass body (upper-middle panel), the FG-to-SG ratio of binary fractions (lower-middle panel) and the FG-to-SG ratio of velocity dispersion with binaries as separated stars (lower panel), respectively for the \srC{}{x3-800} model. Different colors and line styles represent different times. The projected radius is normalized by the half-mass radius in projection. 50 snapshots are combined for better statistics. 
 }\label{F7}
\end{figure}
To investigate the energy equipartition process, we measure the slopes of velocity dispersion scaling, $\sigma(m)\propto m^{\eta}$. Although we do not consider any realistic mass function but assume that all stars have the same mass, for this analysis binaries can be regarded as single stars with positions equal to their center-of-mass  positions and masses equal to 2$m_{\rm s}$.
Fig. \ref{F5} shows the time evolution of $\eta$ within the 5\% Lagrangian radius ($\eta_{5\%}$) and within the half-mass radius ($\eta_{\rm h}$) for all models. 
The evolution of $\eta$ for the inert binary model is also plotted for comparison; $\eta$ rapidly decreases with time until the cluster's core contraction is halted by the supply of energy provided by binary stars.
In general, $\eta_{5\%}$ is smaller than $\eta_{\rm h}$ as previously noted by several studies \citep[i.e. the inner regions can become closer to the equipartition than the outer regions, see e.g.][]{2013MNRAS.435.3272T,2017MNRAS.464.1977W}.

It is interesting to point out that, as shown in Fig. \ref{F5}, models with primordial binary stars can approach full energy equipartition in their central regions more closely than the model with inert binaries; this is due to the the continuous 'supply' of single stars brought in the inner regions as binary members released as single stars due to binary disruption. This combination of binary segregation and disruption provides a mechanism to populate the cluster's inner regions with single stars where they can interact with binaries and cause the cluster to continue to evolve towards the energy equipartition. Without such a mechanism, the inner regions of the inert binary model become dominated by massive stars (the inert binaries) and the evolution of $\eta_{5\%}$ stops earlier when the cluster is less close to a state of full energy equipartition.

Finally, Fig. \ref{F6} shows the time evolution of $\eta$ for SG and FG stars separately; we do not find significant differences between the evolution of $\eta$ for FG and SG stars.

\subsection{Implications of binary spatial mixing for the cluster's kinematics}
In this section we explore the possible implications of the differences between the fraction and the spatial distributions of FG and SG binaries for the kinematic properties inferred from line-of-sight (LOS) velocity measurements. Unresolved binaries can play an important role in shaping the velocity dispersion profiles of clusters regardless of their small fraction since their internal motion can inflate significantly the LOS velocity dispersion. 

Recently, \citet{2018ApJ...864...33D} found that in NGC 6362 the LOS velocity dispersion of SG stars is systematically smaller than that of FG stars. This kinematic difference is puzzling since  a previous study \citep{2014ApJ...791L...4D} has shown that the two different populations of this cluster are spatially mixed.

The analysis of \citet{2018ApJ...864...33D} shows that the observed difference between the kinematic properties of FG and SG stars in NGC 6362 could be due to an SG binary fraction lower than that of the FG population.
The observations of \citet{2018ApJ...864...33D} show that indeed the binary fraction of the SG population ($\sim$1 per cent) is smaller than that of FG stars ($\sim$15 per cent) which is also consistent, in general, with previous theoretical \citep[e.g.][\hyperlink{P1}{Papers I} and \hyperlink{P2}{II}]{2011MNRAS.416..355V} and observational \citep{2015A&A...584A..52L} studies.

The radial profile of the ratio of the FG to the SG velocity dispersions found in \citet{2018ApJ...864...33D} is approximately equal to one in the inner regions and increases as the distance from the cluster's centre increases. The observed radial variation of the ratio of the FG and SG velocity dispersions is particularly interesting and is consistent with the expected radial variation of the FG and SG binary incidence that we found in \hyperlink{P2}{Paper II}; there is no significant difference between the FG and SG binary fractions in the central regions but this difference increases at larger distances from the cluster's centre. Fig. \ref{F7} shows the time evolution of the spatial distribution and velocity dispersion radial profile for FG and SG stars for a representative model, \srC{}{x3-800}. The evolution towards spatial and kinematic mixing is illustrated by the top two panels, which show the time evolution of the radial profiles of the SG-to-FG number ratio and the ratio of the FG-to-SG LOS velocity dispersion  in which binaries are included using the center-of-mass position and velocity.

The bottom two panels of Fig. \ref{F7} show the evolution of the radial profile of the ratio of the FG to the SG binary fraction and of the ratio of the FG to the SG stars' LOS velocity dispersion where velocity dispersions are in this case measured by including the internal motion of individual components of binary stars. The evolution of the ratio of the FG to the SG binary fraction further illustrates the preferential SG binary disruption and the increasing FG binary fraction (relative to the SG binary fraction) at larger distances from the cluster's centre (see also \hyperlink{P2}{Paper II}). The delay in complete mixing in the binary stars discussed in the previous sections leads to a radial variation in the fraction of FG and SG binaries. The radial increase of the ratio of the FG to the SG binary fraction, in turn, implies that the effect of the velocity dispersion inflation due to the individual components of binary stars is larger for the FG population at larger distances from the cluster's centre, which provides a kinematic fingerprint of the differences in the dynamical evolution, survival and mixing of FG and SG binaries. The radial variation in the ratio of the LOS velocity dispersion we find is qualitatively consistent with that revealed by \citet{2018ApJ...864...33D}.

\section{Summary and conclusion}
Spatial mixing of multiple stellar populations of globular clusters in which stars formed with different initial spatial distributions is a natural consequence of both internal (i.e., relaxation-driven diffusion) and external (i.e., tidal stripping) dynamical effects \citep{2013MNRAS.429.1913V,2015MNRAS.454.2166M}.
In this paper we have studied the process of mixing for multiple population binary stars.
In our previous studies \citep{2011MNRAS.416..355V,2015MNRAS.449..629H,2016MNRAS.457.4507H},  we have shown that the dynamical processes determining the evolution and survival of binaries (such as disruption, hardening and ejection) can preferentially affect the initially more concentrated SG binary population and lead to significant differences in the fraction and internal orbital properties of SG and FG binaries.

In this study, we have shown that the dynamical processes affecting the evolution of binary stars have an impact also on the evolution of their spatial distribution and the spatial mixing of FG and SG binaries.
Our simulations illustrate the dependence of the binary evolution toward mixing on the internal binding energy and the consequences of the different dynamical timescales (segregation, binary hardening and disruption) of populations with different initial spatial distributions.

The evolution toward spatial mixing is affected by the following processes:\\
1) segregation of binary stars towards the cluster's inner regions; this process is in general more rapid for the initially more concentrated SG stars and has the effect, at least initially, to further increase the differences between spatial distributions of FG and SG binary stars (see Figs. \ref{F1} and \ref{F2});\\
2) binary disruption; for soft binaries the effect of segregation on the spatial differences between FG and SG binaries can be efficiently counteracted by the rapid disruption of binaries as they migrate toward the cluster's inner regions (see the top left panels of Figs. \ref{F1} and \ref{F2}) and result in an evolution similar to that of single stars although the underlying dynamics is different and driven by binary disruption in the inner regions (see also top row of Fig. \ref{F3});\\
3) binary hardening and ejection; the effects of binary disruption are relevant mainly for soft binaries. Although hard binaries are not disrupted during single-binary encounters, they can still be disrupted during binary-binary encounters or ejected after binary-single and binary-binary encounters. As shown in our simulations, these processes can not initially counteract the effect of segregation and the differences in the spatial distributions of SG and FG binaries increases. As the cluster evolves and its central density increases, it eventually reaches a stage in which binary interactions halt the core's contraction. At this point, the inner regions of the SG binary subsystem stops evolving because of the approximate balance between the replenishment of binaries by segregation from the outer regions and the disruption and ejection due to binary-binary and binary-single interactions (see Fig. \ref{F4}). At this point the differences between the SG and FG spatial distributions starts to decrease and the two binary population mix as FG binaries segregate toward the inner regions. The resulting spatial mixing timescale for binaries is, however, longer than that of single stars. Mixing for hard binaries is therefore driven by the interplay between segregation, ejection, and disruption.

The interplay between binary segregation, disruption and ejection determines also the evolution of the mean mass of stars in a cluster's inner regions: the rapid disruption of soft binaries prevents the increase of the mean mass in the cluster's inner regions while as the hardness of the binary population increases so does the fraction of binaries segregating and surviving in the cluster's inner regions leading to a increase of the inner mean mass (see Fig. \ref{F4}). The continuous disruption of binary stars in the inner regions of the cluster and the ensuing 'supply' of single stars from disrupted binaries implies that in the presence of primordial binaries the cluster's inner regions can approach full energy equipartition more closely than a cluster's without a significant fraction of binary stars (see Figs. \ref{F5} and \ref{F6}).

Finally, we have explored the kinematic implications of the binary mixing process and the role played by the different dynamical ingredients affecting this process. The longer timescale associated to the FG-SG binary spatial mixing implies that, even when single stars are already completely mixed, a cluster can still be characterized by the presence of a radial gradient in the fraction of FG and SG binaries with a ratio of the FG to the SG binary fraction increasing with the distance from the cluster's centre. While a direct observation of this radial gradient might be difficult, we have shown that a fingerprint of this radial variation may be present in the radial profile of the ratio of the FG to the SG line-of-sight (LOS) velocity dispersions, $\sigma_{\rm FG}/\sigma_{\rm SG}$. The LOS velocity dispersion inflation due to binary stars affects the FG population more than the SG population and is more significant at larger distances from the cluster's center leading a $\sigma_{\rm FG}/\sigma_{\rm SG}$ increasing with the distance from cluster's centre (see Fig. \ref{F7}). A recent observational study \citep{2018ApJ...864...33D} of the kinematics of multiple populations in NGC 6362 has provided possible evidence of this effect but additional kinematic studies are necessary to seek further evidence of the interesting interplay between a cluster's internal dynamics, multiple populations spatial mixing and the dynamical effects on the evolution and survival of binary stars.

\section*{Acknowledgement}
We thank Nathan Leigh for useful comments that helped to improve the paper.
JH acknowledges support from the China Postdoctoral Science Foundation, Grant No. 2017M610694.
This research was supported in part by Lilly Endowment, Inc.,
through its support for the Indiana University Pervasive Technology
Institute, and in part by the Indiana METACyt Initiative. The
Indiana METACyt Initiative at IU is also supported in part by
Lilly Endowment, Inc.


\begin{thebibliography}{}
\bibitem[Aarseth(2003)]{2003gnbs.book.....A} Aarseth S.~J.\ 2003, Gravitational N-Body Simulations, Cambridge University Press, 430 pp.
\bibitem[Bastian et al.(2013)]{2013MNRAS.436.2398B} Bastian N. et al.,\ 2013, \mnras, 436, 2398 
\bibitem[\protect\citeauthoryear{Bekki}{2011}]{2011MNRAS.412.2241B} Bekki K., 2011, MNRAS, 412, 2241 
\bibitem[Bellini et al.(2009)]{2009A&A...507.1393B} Bellini A. et al.,\ 2009, \aap, 507, 1393 
\bibitem[Bellini et al.(2015)]{2015ApJ...810L..13B} Bellini A. et al.,\ 2015, \apjl, 810, L13 
\bibitem[\protect\citeauthoryear{Breen}{2018}]{2018MNRAS.tmpL.171B} Breen P.~G., 2018, MNRAS, 
\bibitem[\protect\citeauthoryear{Chatterjee et al.}{2010}]{2010ApJ...719..915C} Chatterjee S., Fregeau J.~M., Umbreit S., Rasio F.~A., 2010, ApJ, 719, 915 
\bibitem[\protect\citeauthoryear{Chatterjee et al.}{2013}]{2013ApJ...777..106C} Chatterjee S., Rasio F.~A., Sills A., Glebbeek E., 2013, ApJ, 777, 106 
\bibitem[Cordero et al.(2014)]{2014ApJ...780...94C} Cordero M.~J. et al.,\ 2014, \apj, 780, 94 
\bibitem[\protect\citeauthoryear{Cordero et al.}{2017}]{2017MNRAS.465.3515C} Cordero M.~J., H{\'e}nault-Brunet V., Pilachowski C.~A., Balbinot E., Johnson C.~I., Varri A.~L., 2017, MNRAS, 465, 3515 
\bibitem[Dalessandro et al.(2014)]{2014ApJ...791L...4D} Dalessandro E. et al.,\ 2014, \apjl, 791, L4 
\bibitem[\protect\citeauthoryear{Dalessandro et al.}{2018a}]{2018ApJ...859...15D} Dalessandro E., et al., 2018a, ApJ, 859, 15 
\bibitem[\protect\citeauthoryear{Dalessandro et al.}{2018b}]{2018ApJ...864...33D} Dalessandro E., et al., 2018b, ApJ, 864, 33 
\bibitem[\protect\citeauthoryear{D'Antona et al.}{2016}]{2016MNRAS.458.2122D} D'Antona F., Vesperini E., D'Ercole A., Ventura P., Milone A.~P., Marino A.~F., Tailo M., 2016, MNRAS, 458, 2122 
\bibitem[Decressin et al.(2007)]{2007A&A...464.1029D} Decressin T., Meynet G., Charbonnel C., Prantzos N., Ekstr{\"o}m, S.\ 2007, \aap, 464, 1029
\bibitem[\protect\citeauthoryear{de Grijs et al.}{2013}]{2013ApJ...765....4D} de Grijs R., Li C., Zheng Y., Deng L., Hu Y., Kouwenhoven M.~B.~N., Wicker J.~E., 2013, ApJ, 765, 4 
\bibitem[\protect\citeauthoryear{de Mink et al.}{2009}]{2009A&A...507L...1D} de Mink S.~E., Pols O.~R., Langer N., Izzard R.~G., 2009, A\&A, 507, L1 
\bibitem[D'Ercole et al.(2008)]{2008MNRAS.391..825D} D'Ercole A., Vesperini E., D'Antona F., McMillan S.~L.~W., Recchi S.\ 2008, \mnras, 391, 825 
\bibitem[\protect\citeauthoryear{D'Ercole et al.}{2012}]{2012MNRAS.423.1521D} D'Ercole A., D'Antona F., Carini R., Vesperini E., Ventura P., 2012, MNRAS, 423, 1521 
\bibitem[\protect\citeauthoryear{D'Orazi et al.}{2010}]{2010ApJ...719L.213D} D'Orazi V., Gratton R., Lucatello S., Carretta E., Bragaglia A., Marino A.~F., 2010, ApJ, 719, L213 
\bibitem[\protect\citeauthoryear{Fare, Webb, \& Sills}{2018}]{2018MNRAS.481.3027F} Fare A., Webb J.~J., Sills A., 2018, MNRAS, 481, 3027 
\bibitem[\protect\citeauthoryear{Ferraro et al.}{2012}]{2012Natur.492..393F} Ferraro F.~R., et al., 2012, Natur, 492, 393 
\bibitem[\protect\citeauthoryear{Geller et al.}{2013}]{2013ApJ...779...30G} Geller A.~M., de Grijs R., Li C., Hurley J.~R., 2013, ApJ, 779, 30 
\newpage
\bibitem[Geller et al.(2015)]{2015ApJ...805...11G} Geller A.~M., de Grijs R., Li C., Hurley J.~R.\ 2015, \apj, 805, 11 
\bibitem[Giersz \& Heggie(1996)]{1996MNRAS.279.1037G} Giersz M., Heggie D.~C.\ 1996, \mnras, 279, 1037 
\bibitem[\protect\citeauthoryear{Gieles et al.}{2018}]{2018MNRAS.478.2461G} Gieles M., et al., 2018, MNRAS, 478, 2461 
\bibitem[Gratton, Carretta \& Bragaglia(2012)]{2012A&ARv..20...50G} Gratton R.~G., Carretta E., Bragaglia A.\ 2012, \aapr, 20, 50 
\bibitem[Heggie \& Hut(2003)]{2003gmbp.book.....H} Heggie D., Hut P.\ 2003, The Gravitational Million-Body Problem: A Multidisciplinary Approach to Star Cluster Dynamics, Cambridge University Press, 372 pp.
\bibitem[Heggie, Trenti \& Hut(2006)]{2006MNRAS.368..677H} Heggie, D.~C., Trenti, M., Hut, P.\ 2006, \mnras, 368, 677 
\bibitem[\protect\citeauthoryear{H{\'e}nault-Brunet et al.}{2015}]{2015MNRAS.450.1164H} H{\'e}nault-Brunet V., Gieles M., Agertz O., Read J.~I., 2015, MNRAS, 450, 1164 
\bibitem[Hong et al.(2013)]{2013MNRAS.430.2960H} Hong J., Kim E., Lee H.~M., Spurzem R.\ 2013, \mnras, 430, 2960 
\bibitem[Hong et al.(2015)]{2015MNRAS.449..629H} Hong J., Vesperini E., Sollima A., McMillan S. L. W., D'Antona F., D'Ercole
A.,\ 2015, \mnras, 449, 629 \hypertarget{P1}{(Paper I)}
\bibitem[Hong et al.(2016)]{2016MNRAS.457.4507H} Hong J., Vesperini E., Sollima A., McMillan S. L. W., D'Antona F., D'Ercole
A.,\ 2016, \mnras, 457, 4507 \hypertarget{P2}{(Paper II)}
\bibitem[\protect\citeauthoryear{Hong et al.}{2017}]{2017MNRAS.464.2511H} Hong J., Vesperini E., Belloni D., Giersz M., 2017, MNRAS, 464, 2511 
\bibitem[\protect\citeauthoryear{Hui, Cheng, \& Taam}{2010}]{2010ApJ...714.1149H} Hui C.~Y., Cheng K.~S., Taam R.~E., 2010, ApJ, 714, 1149 
\bibitem[Hurley et al.(2007)]{2007ApJ...665..707H} Hurley, J.~R., Aarseth, S.~J., \& Shara, M.~M.\ 2007, \apj, 665, 707 
\bibitem[\protect\citeauthoryear{Hypki \& Giersz}{2017}]{2017MNRAS.466..320H} Hypki A., Giersz M., 2017, MNRAS, 466, 320 
\bibitem[Khalisi, Amaro-Seoane \& Spurzem(2007)]{2007MNRAS.374..703K} Khalisi E., Amaro-Seoane P., Spurzem R.\ 2007, \mnras, 374, 703 
\bibitem[King(1966)]{1966AJ.....71...64K} King I.~R.\ 1966, \aj, 71, 64 
\bibitem[Lardo et al.(2011)]{2011A&A...525A.114L} Lardo C. et al.,\ 2011, \aap, 525, A114
\bibitem[Lee(2017)]{2017ApJ...844...77L} Lee J.-W.\ 2017, \apj, 844, 77 
\bibitem[\protect\citeauthoryear{Leigh, Sills, \& Knigge}{2011}]{2011MNRAS.416.1410L} Leigh N., Sills A., Knigge C., 2011, MNRAS, 416, 1410 
\bibitem[\protect\citeauthoryear{Leigh et al.}{2013}]{2013MNRAS.428..897L} Leigh N., Knigge C., Sills A., Perets H.~B., Sarajedini A., Glebbeek E., 2013, MNRAS, 428, 897 
\bibitem[\protect\citeauthoryear{Leigh et al.}{2015}]{2015MNRAS.446..226L} Leigh N.~W.~C., Giersz M., Marks M., Webb J.~J., Hypki A., Heinke C.~O., Kroupa P., Sills A., 2015, MNRAS, 446, 226 
\bibitem[\protect\citeauthoryear{Li, de Grijs, \& Deng}{2013}]{2013MNRAS.436.1497L} Li C., de Grijs R., Deng L., 2013, MNRAS, 436, 1497
\bibitem[\protect\citeauthoryear{Libralato et al.}{2018}]{2018ApJ...861...99L} Libralato M., et al., 2018, ApJ, 861, 99 
\bibitem[Lucatello et al.(2015)]{2015A&A...584A..52L} Lucatello S. et al.,\ 2015, \aap, 584, A52 
\bibitem[Massari et al.(2016)]{2016MNRAS.458.4162M} Massari D. et al.,\ 2016, \mnras, 458, 4162 
\bibitem[Miholics, Webb \& Sills(2015)]{2015MNRAS.454.2166M} Miholics M., Webb J.~J., Sills A.\ 2015, \mnras, 454, 2166 
\bibitem[Milone et al.(2012)]{2012ApJ...744...58M} Milone A.~P. et al.,\ 2012, \apj, 744, 58 
\bibitem[Milone et al.(2013)]{2013A&A...555A.143M} Milone A.~P. et al.,\ 2013, \aap, 555, A143 
\bibitem[\protect\citeauthoryear{Milone et al.}{2017}]{2017MNRAS.464.3636M} Milone A.~P., et al., 2017, MNRAS, 464, 3636 
\bibitem[Milone et al.(2018)]{2018MNRAS.479.5005M} Milone, A.~P., Marino, A.~F., Mastrobuono-Battisti, A., \& Lagioia, E.~P.\ 2018, \mnras, 479, 5005
\bibitem[\protect\citeauthoryear{Nardiello et al.}{2018}]{2018MNRAS.477.2004N} Nardiello D., et al., 2018, MNRAS, 477, 2004 
\bibitem[Nitadori \& Aarseth(2012)]{2012MNRAS.424..545N} Nitadori K., Aarseth S.~J.\ 2012, \mnras, 424, 545 
\bibitem[Piotto et al.(2015)]{2015AJ....149...91P} Piotto G. et al.,\ 2015, \aj, 149, 91 
\bibitem[\protect\citeauthoryear{Pooley \& Hut}{2006}]{2006ApJ...646L.143P} Pooley D., Hut P., 2006, ApJ, 646, L143 
\bibitem[\protect\citeauthoryear{Pooley et al.}{2003}]{2003ApJ...591L.131P} Pooley D., et al., 2003, ApJ, 591, L131 
\bibitem[Richer et al.(2013)]{2013ApJ...771L..15R} Richer H.~B. et al.,\ 2013, \apjl, 771, L15 
\bibitem[Simioni et al.(2016)]{2016MNRAS.463..449S} Simioni M. et al.,\ 2016, \mnras, 463, 449 
\bibitem[Sollima et al.(2007)]{2007ApJ...654..915S} Sollima A. et al.,\ 2007, \apj, 654, 915 
\bibitem[Trenti et al.(2007)]{2007MNRAS.374..344T} Trenti, M., Heggie, D.~C., \& Hut, P.\ 2007, \mnras, 374, 344
\bibitem[Trenti \& van der Marel(2013)]{2013MNRAS.435.3272T} Trenti M., van der Marel R.\ 2013, \mnras, 435, 3272 
\bibitem[Ventura et al.(2001)]{2001ApJ...550L..65V} Ventura P., D'Antona F., Mazzitelli I., Gratton R.\ 2001, \apjl, 550, L65 
\bibitem[Vesperini et al.(2011)]{2011MNRAS.416..355V} Vesperini E., McMillan S.~L.~W., D'Antona F., D'Ercole A.\ 2011, \mnras, 416, 355 
\bibitem[Vesperini et al.(2013)]{2013MNRAS.429.1913V} Vesperini E., McMillan S.~L.~W., D'Antona F., D'Ercole A.\ 2013, \mnras, 429, 1913
\bibitem[Vesperini et al.(2018)]{2018MNRAS.476.2731V} Vesperini E., Hong J., Webb J.~J., D'Antona F., D'Ercole A.\ 2018, \mnras, 476, 2731 
\bibitem[Webb \& Vesperini(2017)]{2017MNRAS.464.1977W} Webb J.~J., Vesperini E.\ 2017, \mnras, 464, 1977 

\end{thebibliography}
\end{document}